# Tunable Light Filtering in a Bragg Mirror/Heavily-Doped Semiconducting Nanocrystal Composite

*Ilka Kriegel[1], Francesco Scotognella[1,2]\**

[1] Dipartimento di Fisica, Politecnico di Milano, Piazza Leonardo da Vinci 32, 20133 Milano, Italy

[2] Center for Nano Science and Technology@PoliMi, Istituto Italiano di Tecnologia, Via Giovanni Pascoli 70/3, 20133 Milano

ABSTRACT

Tunable light filters are critical components for many optical applications in which light in-coupling, out-coupling or rejection is crucial, such as lasing, sensing, photovoltaics and information and communication technology. For this purpose, Bragg mirrors, band-pass filters with high reflectivity represent good candidates. However, their optical characteristics are determined at the stage of fabrication. Heavily doped semiconductor nanocrystals (NCs) on the other hand deliver a high degree of optical tunability through the active modulation of their carrier density ultimately influencing their plasmonic absorption properties. Here, we propose the design of a tunable light filter composed of a Bragg mirror and a layer of plasmonic semiconductor NCs. We demonstrate that the filtering properties of the coupled device can be tuned to cover a wide range of frequencies from the visible to the near infrared (vis-NIR) spectral region when employing varying carrier densities. As tunable component we implemented a dispersion of copper selenide ($Cu_{2-x}Se$) NCs and a film of indium tin oxide (ITO) NCs, which are known to show optical tunablility by chemical or electrochemical treatments. We utilized the Mie theory to describe the carrier dependent plasmonic properties of the $Cu_{2-x}Se$ NC dispersion and the effective medium theory to describe the optical characteristics of the ITO film. The transmission properties of the Bragg mirror have been modelled with the transfer matrix method. We foresee ease of experimental realization of the coupled device, where filtering modulation is achieved upon chemical and electrochemical post-fabrication treatment of the heavily doped semiconductor NC component, eventually resulting in tunable transmission properties of the coupled device.

**Introduction**
Optical filters are fundamental components for almost the totality of the optical setups and devices. For example, they play a very important role in microfluidic devices that are very interesting for point-of-care diagnostics[1,2]. Very interesting strategies to fabricate colour filters, also without auto-fluorescence, are proposed in literature[3]. Although, in many applications, as microfluidic devices that imply the detection of more than one analytes, the use of tunable filters would be a great advantage.

Very efficient filters are Bragg mirrors, also called one-dimensional photonic crystals[4-6]. Bragg mirrors can be made of dense materials[7], but also with mesoporous materials or nanoparticles[8-10]. Their use has been exploited in several types of devices, such as distributed feedback lasers[11-15], sensors[16,17], absorption enhancement for photovolaitcs[18] or in dye sensitized solar cells[19-21]. Furthermore, nanoparticle based photonic crystals have been employed for switching applications[22-25]. An advantage of Bragg mirrors is that a proper design enables to access a variety of different wavelength regions, determined by the design of the respective photonic structure[4-6]. Parameters that can be varied are layer thickness, material refractive index and stacking sequence. However, the tunability is limited to the stage of production.

On a different front, heavily doped semiconductor NCs have been receiving an increasing attention in recent years. Their high level of doping leads to localized surface plasmon resonances (LSPRs) mostly located in the vis-NIR[19]. The advantage of doped semiconductor nanostructures is the option to chemically and electrochemically modify their plasmon resonance frequencies by changing the material's carrier density. For copper chalcogenide NCs, chemical manipulation has been demonstrated in response to oxidizing and reducing treatments[20-24]. The exposure of a solution of copper chalcogenide NCs, such as $Cu_{2-x}S$ or $Cu_{2-x}Se$, to oxygen or the addition of an oxidizing agent leads to a blue-shift and increase in intensity of the NIR LSPR. Notably, this process can be entirely reversed through the addition of reducing agents completing the full cycle of tunability[22-25]. In metal oxide NCs a dynamic modulation of the LSPR has been demonstrated through a fully reversible electrochemical treatment. In this approach a switching of plasmon absorption is achieved through electrochemical doping, activated by applying a voltage to conducting films of metal oxide NCs[26]. In a recent study the modulation of transmittance was achieved electrochemically in layers of transparent conducting NCs such as ITO[26,29]. Moreover, plasmonic effects have been exploited for electrochromic application for the direct modulation of the transmittance of solar energy[27].

In this study, we propose a tunable filter based on photonic crystals, i.e. Bragg mirrors, coupled to heavily doped semiconductor NCs with plasmonic absorption properties. We report the modelling of a Bragg mirror coupled to a dispersion of chemically tunable $Cu_{2-x}Se$ NCs and electrochemically switchable NC films of ITO. In both cases switching is based on a modulation of their carrier density and, thus, this combination will ultimately result in a tunable broadband light transmission. We model the structure by implementing the transfer matrix method, to describe the optical properties of the Bragg mirror. The Mie theory describes the tunable plasmonic properties of the $Cu_{2-x}Se$ NCs, and the effective medium theory to describe the tunable optical characteristics of the ITO film. Since the plasmon peak of the NCs can be dynamically modified, we envisage a filter with properties that can be fine-tuned through a broad range of frequencies according to the desired application.

**Theoretical Methods**

To calculate the transmission spectra of the photonic crystal component in the structure, we have employed the transfer matrix method, a general technique that is widely used in optics for the description of stacked layers and it is extensively described in Ref. 38. We have considered isotropic, nonmagnetic materials shaping the system glass/multilayer/air (in which glass is the sample substrate) and an incidence of the light normal to the stacked layer surface. $n_0$ and $n_S$ are the refractive indexes of air and glass, respectively, while $E_m$ and $H_m$ are the electric and magnetic fields in the glass substrate. To determine the electric and magnetic fields in air, $E_0$ and $H_0$, we have solved the following system:

$$\begin{bmatrix} E_0 \\ H_0 \end{bmatrix} = M_1 \cdot M_2 \cdot .... \cdot M_m \begin{bmatrix} E_m \\ H_m \end{bmatrix} = \begin{bmatrix} m_{11} & m_{12} \\ m_{21} & m_{22} \end{bmatrix} \begin{bmatrix} E_m \\ H_m \end{bmatrix} \quad (1)$$

where

$$M_j = \begin{bmatrix} A_j & B_j \\ C_j & D_j \end{bmatrix},$$

with $j=(1,2,…,m)$, is the characteristic matrix of each layer. The elements of the transmission matrix *ABCD* are

$$A_j = D_j = \cos(\phi_j), \quad B_j = -\frac{i}{p_j}\sin(\phi_j), \quad C_j = -i p_j \sin(\phi_j), \quad (2)$$

where $n_j$ and $d_j$, hidden in the angle $\phi_j$, are respectively the effective refractive index and the thickness of the layer *j*. In the case of normal incidence of the probe beam, the phase variation of the wave passing the *j*-fold layer is $\phi_j = (2\pi/\lambda)n_j d_j$, while the coefficient $p_j = \sqrt{\varepsilon_j/\mu_j}$ in transvers electric wave and $q_j=1/p_j$ replace $p_j$ in transvers magnetic wave. Inserting Equation (4) into Equation (3) and using the definition of the transmission coefficient

$$t = \frac{2p_s}{(m_{11} + m_{12}p_0)p_s + (m_{21} + m_{22}p_0)} \quad (3)$$

it is possible to write the light transmission as

$$T = \frac{p_0}{p_s}|t|^2 \quad (4)$$

where $p_s$ represents the substrate and $p_0$ air. To calculate the optical properties of the doped $Cu_{2-x}Se$ NC dispersion in the region of plasmonic absorption we used the quasi static approximation of Mie scattering theory, which provides for the absorption cross-section $\sigma_A$ the following expression

$$\sigma_A = 4\pi k R^3 \operatorname{Im}\left(\frac{\varepsilon(\omega) - \varepsilon_m}{\varepsilon(\omega) + 2\varepsilon_m}\right) \quad (5)$$

where $\varepsilon(\omega)$ is the bulk dielectric function at the optical frequency $\omega$ and $\varepsilon_m$ is the dielectric constant of the surrounding medium. The optical extinction of ultra-small particles is dominated by absorption, but a first order correction to quasi-static approximation ought to be included to account for a scattering contribution, with a scattering cross-section given by the following expression

$$\sigma_S = \frac{8}{3}\pi k^4 R^6 \left| \frac{\varepsilon(\omega) - \varepsilon_m}{\varepsilon(\omega) + 2\varepsilon_m} \right|^2 \quad (6)$$

In the above equations, $k = n_m \omega / c$ with $n_m = (\varepsilon_m)^{1/2}$ being the refractive index of the dielectric environment and $c$ is the speed of light in vacuum[39]. The Drude model has been assumed to account for the optical properties in heavily doped semiconductor NCs in the NIR, with a complex dielectric function given by[27-31]:

$$\varepsilon(\omega) = \varepsilon_1(\omega) + i\varepsilon_2(\omega) \quad (7)$$

where

$$\varepsilon_1 = \varepsilon_\infty - \frac{\omega_P^2}{\omega^2 + \Gamma^2}$$

$$\varepsilon_2 = \frac{\omega_P^2 \Gamma}{\omega(\omega^2 + \Gamma^2)} \quad (8)$$

with $\Gamma$ the free carrier damping and

$$\omega_P = \sqrt{\frac{N_C e^2}{m^* \varepsilon_0}} \quad (9)$$

the plasma frequency of the free carriers of the system, where $N_C$ the carrier density, $e$ the charge of the electron, $m^*$ the effective mass and $\varepsilon_0$ the vacuum dielectric permittivity.

The absorption of the ITO NC film has been described by applying the effective medium theory or Maxwell-Garnett effective medium approximation (MG-EMA)[40-42]. This theory has been developed to describe the macroscopic properties of a composite material and to average the medium dielectric function according to the multiple values of the constituents of the composite material. The effective dielectric function ($\varepsilon_{eff}$) of a film of ITO NCs can be described by the MG-EMA as follows:

$$\varepsilon_{eff} = \varepsilon_m \frac{2(1-\delta_i)\varepsilon_m + 2(1+\delta_i)\varepsilon_i}{(2+\delta_i)\varepsilon_m + (1-\delta_i)\varepsilon_i} \quad (10)$$

where $\varepsilon_m$ is the medium dielectric constant, $\varepsilon_i$ is the frequency dependent dielectric function of the bulk material (in this case approximated by the Drude model, equations 8 and 9) and $\delta_i$ accounts for the volume fraction. Within this theory only far field interactions are taken into account, while near field interaction among the NCs are neglected. The absorbance of the NC films is finally calculated from the imaginary part of the dielectric function [$(\varepsilon_{eff})^{1/2}$] and the Beer–Lambert law[42].

**Results and Discussion**

In Figure 1 we show a schematic of the proposed device, a Bragg mirror (blue and orange layers in Figure 1) coupled to either a dispersion of $Cu_{2-x}Se$ NCs or an ITO NC film, sketched in dark red.

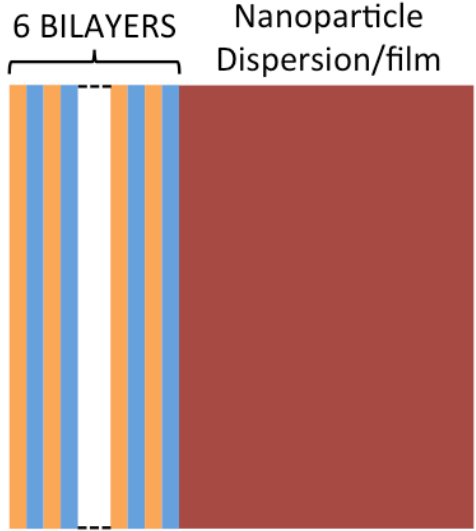

**Figure 1**. Scheme of the device in which a Bragg mirror (photonic crystal), illustrated by alternating orange and blue layers, is coupled to a dispersion of $Cu_{2-x}Se$ NCs in toluene or an ITO NC film (dark red) to act as a tunable light filter. In the presented device tunability is envisaged upon the modulation of the carrier density in the heavily doped semiconductor NC component through chemical and electrochemical treatments, ultimately leading to tunable plasmonic light absorption.

In the following we demonstrate the modelling of the transmission properties of the proposed device. We will first consider the plasmonic absorption properties of the heavily doped NC component and demonstrate its modulated absorption properties through an appropriate choice of carrier densities in the structure. Thereafter, we will demonstrate how the absorption of the Bragg mirror is altered, when coupled to a layer of NCs with varying carrier densities and finally present the transmission properties of the actual device. Concerning the $Cu_{2-x}Se$ NCs, we have considered a dispersion of $Cu_{2-x}Se$ NCs in toluene with spherical shape and diameter of 15 nm, with $\varepsilon_\infty=11$[27, 43]. The carrier density dependent effective mass and damping constant were taken from Ref. 43 with the following parameters: $m_1^*=0.445m_0$, $m_2^*=0.394m_0$, $m_3^*=0.334m_0$, and $m_4^*=0.336m_0$ and $\Gamma_1=0.189$ eV, $\Gamma_2=0.232$ eV, $\Gamma_3=0.244$ eV, and $\Gamma_4=0.254$ eV for carrier densities of 1.14, 1.53, 1.95 and $2.58 \times 10^{21}$ cm$^{-3}$, respectively. The dispersion is in toluene, such that $\varepsilon_m=2.24$. The concentration of the NCs in the dispersion has been chosen to be $1 \times 10^{17}$ cm$^{-3}$, and the dispersion thickness is 1 millimeter. It has been demonstrated that, by oxidation or reduction of the NCs, the carrier density can be pitched, with a consequent variation of the plasma frequency. In Figure 2a we show the calculated absorption spectra of $Cu_{2-x}Se$ NCs for the different carrier densities, as given above and the corresponding parameters of effective mass and damping constant. A blue shift of up to 0.7 eV and increase in intensity with increasing carrier density is observed. The calculated results are in good agreement with experimental results, obtained in Ref. 29-32. In those works, a modulation of the plasmonic absorption has been triggered through the addition of chemical agents inducing oxidation and reduction. This in turn leads to a variation of the carrier density and a blue (for oxidation) or red shift (for reduction) of the plasmon resonance over a wide range of frequencies. Such observation underlines that our model appropriately describes the absorption properties of a $Cu_{2-x}Se$ NC dispersion, where a tuning of the plasmon resonance can be triggered by chemical means.

As a second example we have considered a film of ITO NCs with a thickness of 5 micrometers employing the MG-EMA to calculate its absorption properties[28]. The refractive index of the film is 1.5 (as the one of glass). For the NCs we consider a free carrier damping $\Gamma=0.31$ eV and a high frequency dielectric permittivity of $\varepsilon_\infty=4$[28-30]. The volume fraction of ITO in the film is chosen to be 0.01[35]. In Figure 2b the absorption spectra of the ITO film with different carrier densities, namely 0.7, 1, 1.3 and $2 \times 10^{21}$ cm$^{-3}$ are depicted. This theoretical approach has been successfully employed in a recent work to describe the absorption properties of ITO NC films[37,42]. Experimentally the plasmon frequency of ITO NC films has been reversibly tuned by electrochemical post-treatment leading to a switching of the plasma frequency through the NIR [33,36,37]. Taken altogether the proposed theoretical framework is appropriate to describe the tuning of light absorption with doped semiconductor NCs in dispersion or in films, both thoroughly confirmed in experimental and theoretical works.[29,32,33,36,37]

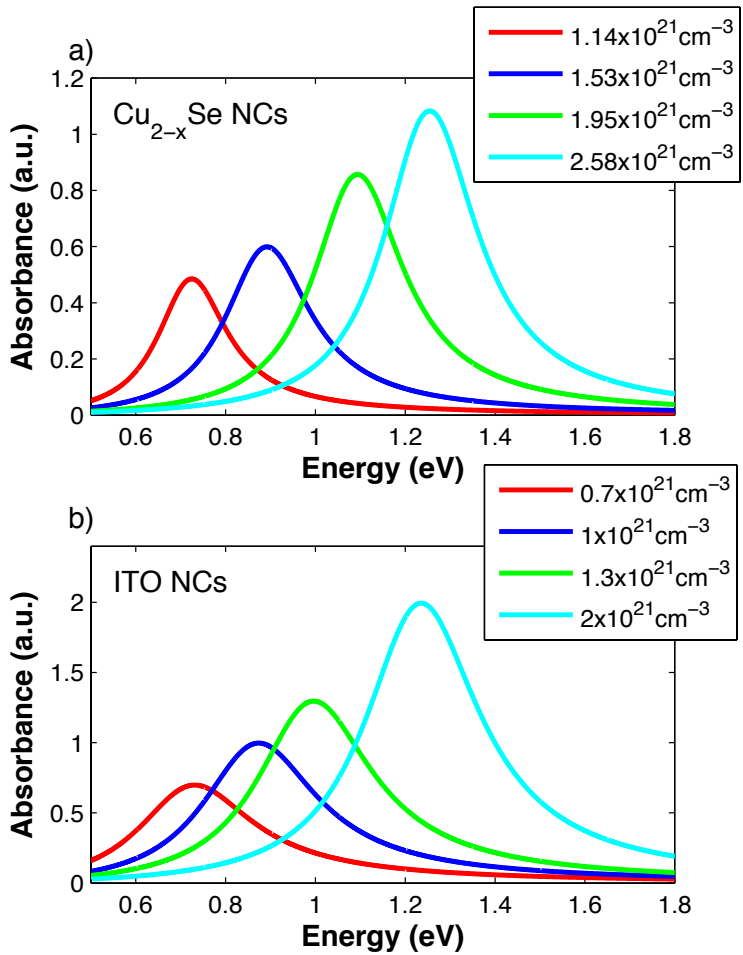

**Figure 2.** Illustration of the tunable optical properties of a) a $Cu_{2-x}Se$ NC dispersion and b) an ITO NC film as a function of their carrier density $N_C$, that are respectively 1.14, 1.53, 1.95 and $2.58 \times 10^{21}$ cm$^{-3}$ and 0.7, 1, 1.3 and $2 \times 10^{21}$ cm$^{-3}$. The given results are calculated according to a) the Mie theory and b) the MG-EMA assuming a Drude like behaviour of the free carriers in the system.

To describe the optical characteristics of the Bragg mirror, we have employed an alternated stack of six bilayers of silicon dioxide ($SiO_2$, with refractive index $n$=1.46) and zinc oxide (ZnO, with refractive index $n$=2) and a layer thickness of 180 nm. We did not consider the refractive index dispersion for these two metal oxides, since they are approximately constant in the range of energies considered in this study. The absorption spectrum of the multilayer shows a peak centred at 1 eV (blue curves in Figure 3), due to the occurrence of the photonic band gap. The chosen materials, layer thicknesses, and number of layers are typical examples of Bragg mirrors. The absorption properties of such stackings are well studied and demonstrated in experimental frameworks, thus, appropriate for the actual applicability of our proposed device[44-45].

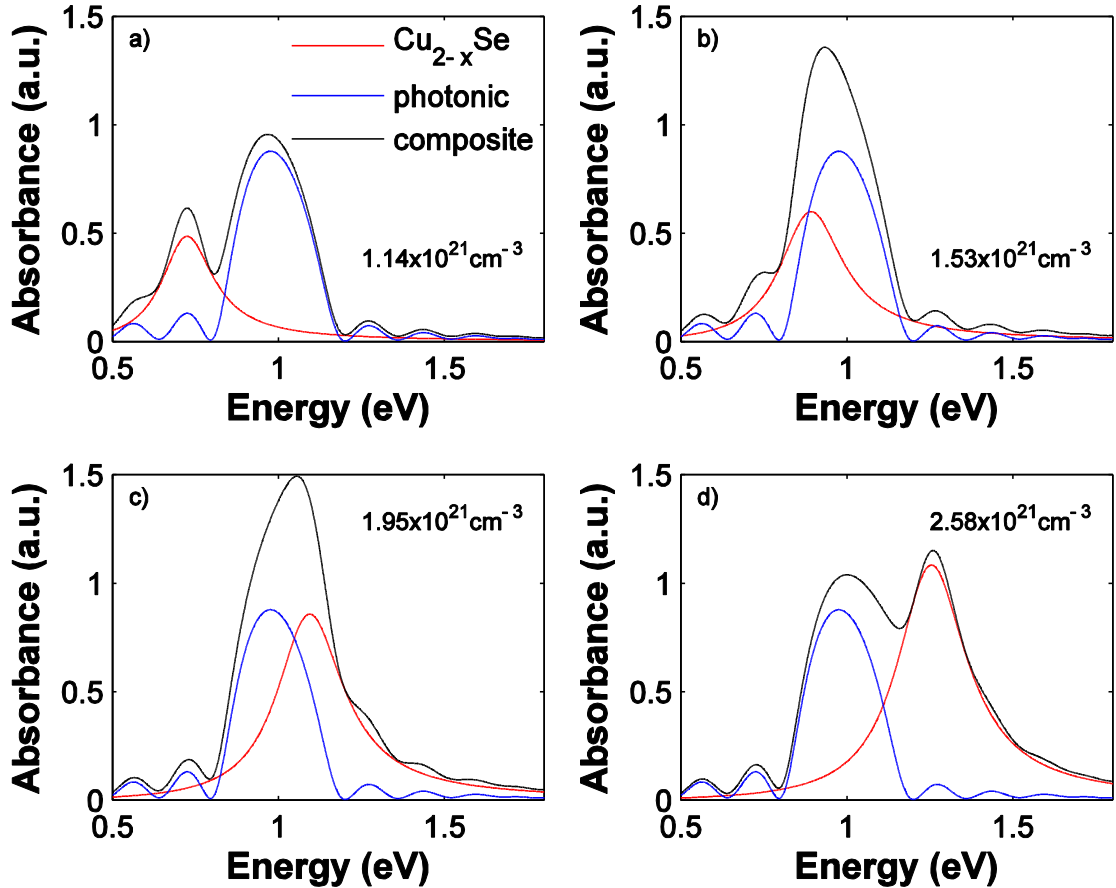

**Figure 3.** Absorption properties of a Bragg mirror (blue curves) coupled to a dispersion of $Cu_{2-x}Se$ NCs (red curves) for different carrier concentrations of 1.14, 1.53, 1.95 and 2.58 × $10^{23}$ $cm^{-3}$ for a), b), c) and d), respectively. The black curves illustrate the optical properties of the coupled device.

By coupling the engineered Bragg mirror with the $Cu_{2-x}Se$ NC dispersion, we obtain a composite device showing an absorption spectrum that is the sum of the photonic band gap and the plasmon absorption (Figure 3). In Figure 3 the blue curve depicts the absorption spectrum of the Bragg mirror alone, characterized by the photonic band gap at around 1 eV. The red curves describe the absorption of the $Cu_{2-x}Se$ NC dispersion for different carrier densities, namely 1.14, 1.53, 1.95 and 2.58 × $10^{23}$ $cm^{-3}$ for Figure 3a, Figure 3b, Figure 3c and Figure 3d, respectively and as calculated above (Figure 2). The black curve finally describes the overall absorption of the Bragg mirror/NC composite. Through the appropriate choice of materials, i.e. $Cu_{2-x}Se$ NC dispersion, and the clever design of the Bragg mirror, the absorption properties of the coupled device result in an extension of the absorption to either the red, for low carrier densities of the heavily doped semiconductor NCs, or the blue for high carrier densities. This combination enables a filtering tunability, which is not achievable for the Bragg mirror alone.

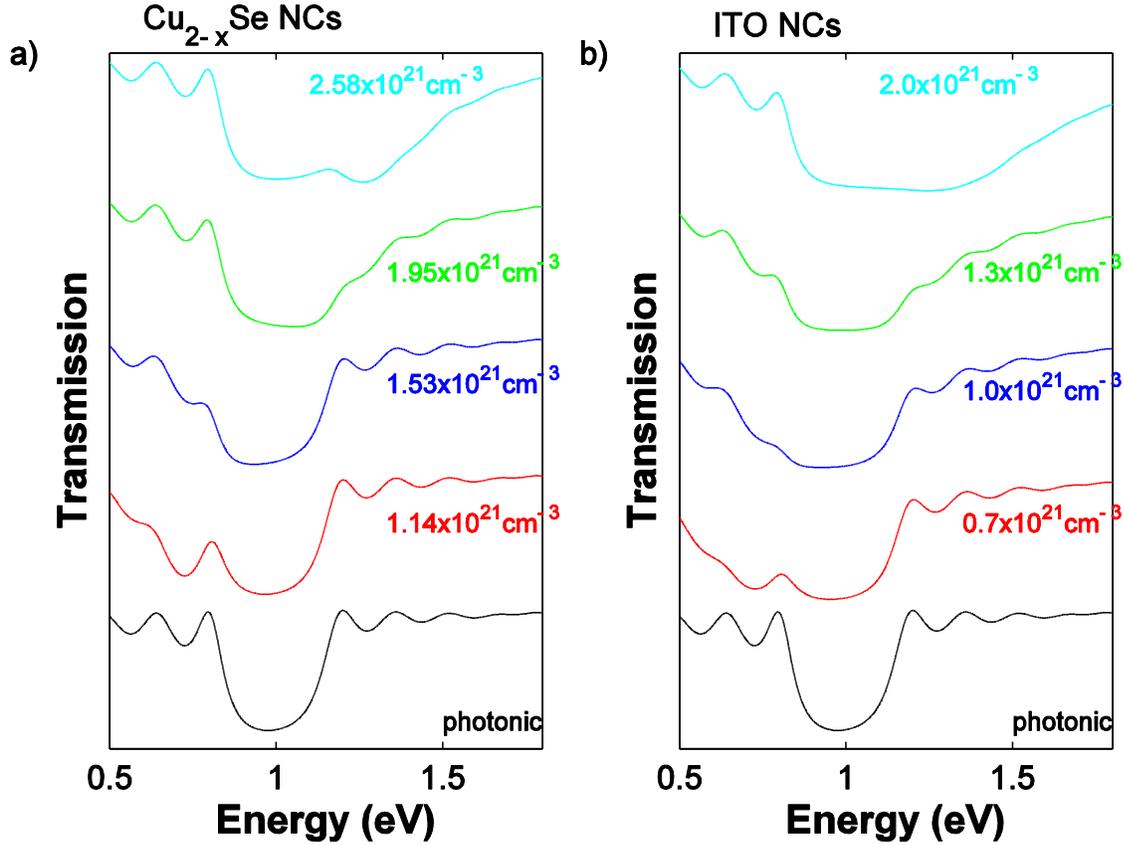

**Figure 4.** Transmission of the tunable light filter for a Bragg mirror coupled to a) a $Cu_{2-x}Se$ NCs dispersion and b) to an ITO NC film for varying carrier concentrations representing the band gap tuning of the Bragg mirror. The black curves correspond to the bare Bragg mirror. In both presented devices the filtering covers a broader range which is blue shifted, for higher carrier concentration and red shifted for lower concentrations, representing the tunability of the filter upon carrier density modulation in the heavily doped semiconductor NC component.

The overall transmission properties of the coupled device are demonstrated in Figure 4, which actually demonstrates its active filtering modulation. We show the transmission spectrum for coupling a Bragg mirror to the $Cu_{2-x}Se$ NC dispersion in Figure 4a and the ITO NC films in Figure 4b. We clearly observe that, with respect to the photonic crystal alone, the composite shows a larger band. Furthermore, by changing the carrier density in the $Cu_{2-x}Se$ NC dispersion and the ITO NC films from 1.14, over 1.53, 1.95 and to $2.58 \times 10^{21}$ cm$^{-3}$, respectively, we observe a coverage of a broader absorption range from the red to the blue part of the visible spectrum. The overall transmission spectrum of the composite is therefore reversibly tunable owing to the possibility to influence dynamically and reversibly the carrier density in the system, inducing a blue or red shift of the plasmonic absorption in the heavily doped NC component. Moreover, by changing the refractive index ratio of the Bragg mirror, the linewidth of its photonic band gap can be varied allowing another degree of freedom in the design of the proposed coupled device (see Supporting Information Figure S1).

The unique combination of a high reflecting Bragg mirror with strongly absorbing NCs and its exceptional tuning properties allows for a filtering device that delivers the outstanding property of active transmission modulation over a broad range of frequencies. We want to point out that we envisage the band gap tuning of the Bragg mirror/NC composite as demonstrated in Figure 4 to be experimentally feasible upon a fully reversible post-fabrication treatment. Once the composite is fabricated by coupling the NC dispersion or film to the Bragg mirror, tuning of the transmission is realized by applying chemical or electrochemical treatments, while keeping the concentration and thickness of the dispersions or films and the Bragg mirror intact. The wavelength tuning is realized solely by influencing the carrier density of the NCs, which in turn leads to the altered absorption properties and the shift through a broad wavelength range, as demonstrated in Figure 4. We further highlight that the transmission modulation in an actual device would not occur in steps, as demonstrated by us through the choice of certain given carrier densities, but would rather change continuously covering the entire range of wavelengths. Moreover, the use of a photonic crystal and of a heavily doped semiconductor NC component provide filtering without autofluorescence [See a detailed discussion of autofluorescence in Ref. 3 and Ref. 46]. For this reason, the proposed filter is particularly appropriate for sensor devices, also for biological applications, where low fluorescence signals are detected.

**Conclusion**

We introduced the design and theoretical framework of a non emissive tunable filter composed of a Bragg mirror coupled to a thin layer of heavily doped, plasmonic semiconductor NCs. The high reflectivity of the Bragg mirror and the strong extinction coefficient of the heavily doped semiconductor NCs, combined with the exceptional tuning properties of the plasmonic absorption results in an effective filtering modulation. We demonstrated that the bandgap tuning over a wide range of frequencies is achieved when the carrier density of the plasmonic NC layer is properly designed. Within this work we envisaged a $Cu_{2-x}Se$ NC dispersion or an ITO NC film as tunable components. In both systems, the experimental modulation of their absorption properties by chemical or electrochemical means has been shown in a number of recent works demonstrating their actual applicability[27-34]. In this work, the calculations carefully take into account all the physical parameters considered to model the aforementioned experimental data. However, we envisage the use of any other type of highly doped semiconductor NC, rendering possible a clever design of the filtering properties. We foresee a straightforward realization of such filters by coupling a NC dispersion or film with a Bragg mirror, deposited with spin coating or sputtering techniques[8,44-45]. The application of this kind of tunable filtering component is very promising for varying of optical applications, such as lasing, sensing, photovoltaics or information and communication technology.

ACKNOWLEDGMENT
I. K. and F.S. acknowledge the project EDONHIST (Grant No. 2012- 0844).

AUTHOR INFORMATION
**\* Corresponding Author**
Francesco Scotognella, Politecnico di Milano, Dipartimento di Fisica, piazza Leonardo da Vinci 32, 20133 Milano, Italy.
Telephone: +39 02 23996056; FAX: +39 02 23996126
Institution email address francesco.scotognella@polimi.it

SUPPORTING INFORMATION

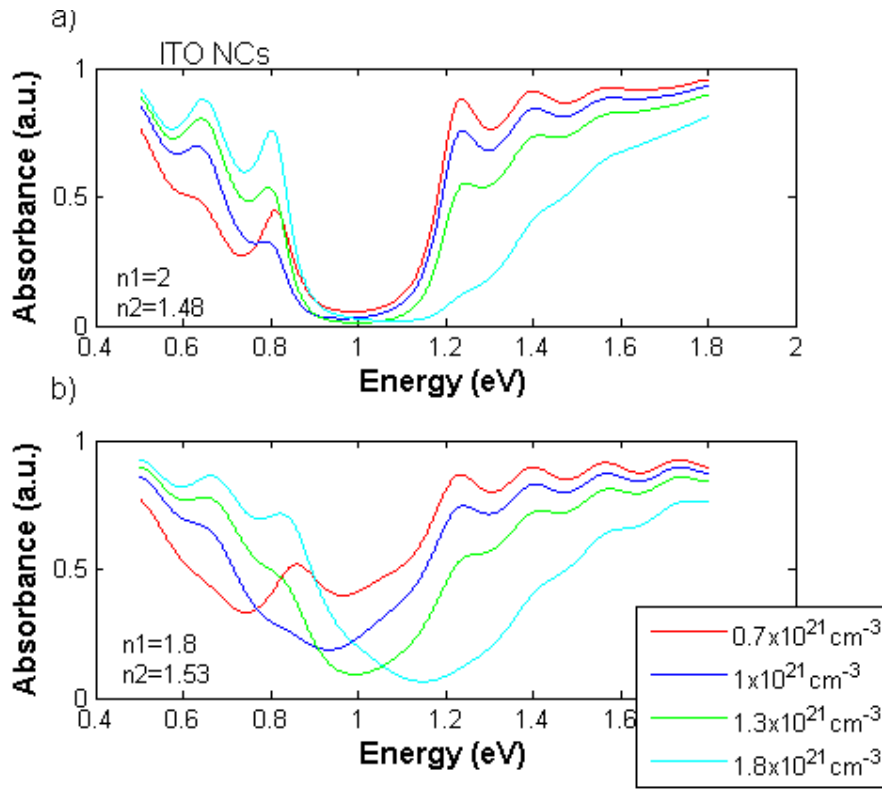

Figure S1 Transmission of the tunable light filter for a Bragg mirror coupled to a) a $Cu_{2-x}Se$ NCs dispersion and b) to an ITO NC film for varying carrier concentrations representing the band gap tuning of the Bragg mirror. The black curves correspond to the bare Bragg mirror. In Figure S1a the refractive indexes of the two materials are 2 and 1.48, respectively, while in Figure S1b the refractive indexes of the two materials are 1.8 and 1.53, respectively.